\documentclass[aps,pra,epsfigure,showpacs,twocolumn]{revtex4}
\usepackage{graphicx}          
\usepackage{epsfig}
\usepackage{amsmath}
\usepackage{latexsym}
\usepackage{amsfonts}
\usepackage{amssymb}
\usepackage{array}

\newcommand{\dd}{\mathrm{d}}
\newcommand{\ee}{\mathrm{e}}
\newcommand{\ii}{\mathrm{i}}

\newcommand{\ket}[1]{\left\vert#1\right\rangle}

\begin{document}
\title{Characterization of the entanglement of two squeezed states}
\author{P. Marek, M. Paternostro\footnote{Present address: Institute for Quantum Optics and Quantum Information (IQOQI), Austrian Academy of Sciences, Boltzmanngasse 3, A-1090 Vienna, Austria} and M. S. Kim}
\affiliation{
School of Mathematics and Physics, Queen's University,
Belfast BT7 1NN, United Kingdom \\
}

\date{\today}

\begin{abstract}

We study a continuous-variable entangled state composed of two
states which are squeezed in two opposite quadratures in phase
space. Various entanglement conditions are tested for the
entangled squeezed state and we study decoherence models for
noise, producing a mixed entangled squeezed state. We briefly
describe a probabilistic protocol for entanglement swapping based
on the use of this class of entangled states and the main features
of a general generation scheme.
\end{abstract}


\maketitle

\section{Introduction}
In these years, we have witnessed an increasing interest related
to the use of continuous variable (CV) systems as appealing
candidates in many applications of quantum information processing
(QIP) and quantum communication~\cite{vanloock}. For instance, the
first experimental observation of CV quantum
teleportation~\cite{furusawa} has been accomplished, by using a CV
entangled channel represented by a simple two-mode squeezed
state~\cite{loudon}, following an earlier CV-based theoretical
proposal~\cite{braunstein}. It is now well-understood that CV
systems embodied by light field modes may serve as reliable,
long-haul information carriers, due to properties of robustness
against decoherence~\cite{carrier}. On the other hand, some
strategies for the performance of quantum computation with CV
systems have been designed in different physical
systems~\cite{computation}.

A key element in these QIP tasks is the generation and
manipulation of non-classical states of CVs embodying
non-classical channels, the paradigmatic examples being
represented by coherent superpositions of macroscopically
distinguishable coherent states~\cite{stoler,jeong} and entangled
coherent states~\cite{sanders1}. However, it is frequently the
case that a full characterization of the non-classical properties
of an entangled state of two CV subsystems is made difficult by
the lack of objective criteria able to discriminate whether or not
a general CV state is entangled. A completely satisfactory answer
to this question is possible, so far, only for the special class
of Gaussian states, {\it i.e.} those CV states whose
characteristic functions are Gaussian~\cite{hwang,barnett}. In
this case, necessary and sufficient criteria for entanglement
exist~\cite{Simon,Duan} which, however, could fail for
non-Gaussian states. In order to bypass this bottleneck, very
recently some alternative strategies have been designed in order
to provide more efficient tests for entanglement in general
bipartite CV states~\cite{Agarwal,Shchukin,Hillery,Paternostro}.
On one hand, the results of this investigation have the merit of
filling the gap between Gaussian and non-Gaussian CV states. On
the other hand, they legitimize the theoretical and practical
consideration of non-Gaussian states in application of QIP and, by
enlarging the class of characterizable CV states, pave the way
toward the study of important aspects of quantum entanglement such
as its resilience to temperature and
mixedness~\cite{iojacobmyung}.

In this work, we perform a further step along the direction of CV
quantum state engineering and characterization by studying a
special type of non-Gaussian CV state, namely the entanglement of
two single-mode squeezed vacuum states, under the QIP perspective.
We address both the cases of pure and mixed states showing that,
through the application of some recently suggested tests (some of
them being intrinsically operative), entanglement can be revealed
regardless of the non-Gaussian nature and the mixedness of the
state considered. The aim of our study is to characterize the
correlation properties of entangled squeezed states (ESS's) and
discuss a possibility of using them in implementation of some
basic quantum protocols, such as teleportation
\cite{teleportation} and entanglement swapping \cite{swapping}. In
order to complete our study, we also sketch two schemes allowing
for the generation of ESS's.

In Section~\ref{entanglement} we introduce the class of ESS's we
consider, and formalize the conditional gate we use in order to
set entanglement between two initially independent single-mode
squeezed states. We then study the properties of these states,
both for the pure and the mixed case, by applying some recently
suggested tests for
entanglement~\cite{Shchukin,Hillery,Paternostro}. We
quantitatively address the case of thermal- and phase-diffusion
noise affecting the purity of the states, quantifying the
entanglement via the use of logarithmic negativity~\cite{logneg}
and, for a simplified impurity model, sketching the results
achieved by applying the entangling power
criterion~\cite{Paternostro}. Section~\ref{teleswap} briefly
addresses the application of ESS's in entanglement swapping and
teleportation protocols, and finally, Section~\ref{scheme}
describes two protocols for the generation of ESS's based on
interactions of CV modes with a two-level system.
Section~\ref{conclusions} summarizes our results.

\section{Characterizing the Entanglement of an ESS}
\label{entanglement}

\subsection{Characterizing the entanglement in a pure ESS}
\label{uno}
A vacuum is a minimum uncertainty state whose mean energy is zero.
In a phase space representation, as it is well-known, it is
possible to modify the uncertainties of the quadrature operators
associated with a particular state, still maintaining its
minimum-uncertainty nature. More in details, by reducing the
uncertainty of one quadrature at the expense of increasing the
uncertainty of the canonically conjugate one, a squeezed vacuum is
produced with a mean energy which becomes a function of the amount
of squeezing. Under a formal point of view, the squeezing
operation is accounted by the unitary transformation~\cite{loudon}
$\hat{S}(s)=\exp\left[{\frac{s}{2}}({a}^{\dag^{2}}-{a}^2)\right]$,
where $s$ is the squeezing parameter and ${a}^\dag$ (${a}$) is the
creation (annihilation) operator of the field. The choice of a
direction, in the phase space, for reducing the noise is
determined by the argument of $s$ which, in general, is a complex
number. For the sake of convenience, without affecting the
generality of our study, here we restrict our attention to the
case of a real $s$. The state resulting from the squeezing of a
initial vacuum state, or a squeezed vacuum, is from now on
indicated as $|\psi_s\rangle=\hat{S}(s)|0\rangle$. Although
squeezed states are now routinely produced, they are usually
burdened by noise which alters their minimum-uncertainty nature
and makes them intrinsically mixed. As such, in general, they need
to be described by a density matrix $\rho_{s+}$. Let us consider a
two-mode entangled squeezed vacuum (ESV) which, assuming the above
conditions of generality, is defined here as
\begin{equation}
\label{ESS}
\rho_{ESV} = \mathcal{T}\left(\otimes_{j=\text{a,b}}\rho_{s+,j}\right)\mathcal{T}^{\dag},
\end{equation}
where $j=a,b$ is a label for the mode considered (mode b being
described by the bosonic operators $b$ and $b^\dag$), each
$\rho_{s+,j}$ is considered as already normalized and ${\cal T}$
is  the unitary transformation
 \begin{equation} \label{Transformation}
{\mathcal{T}} =  \openone_{\text{a}}\otimes{e}^{\ii\frac{\pi}{2}b^{\dag}b}+e^{\ii\phi}e^{\ii\frac{\pi}{2}a^{\dag}a}\otimes\openone_{\text{b}}.
\end{equation}
Here $\openone_j$ is the identity operator for mode $j$, $\phi$ is
an arbitrary relative phase and $e^{\ii\frac{\pi}{2}j^{\dag}j}$
describes a phase-shift (by an amount of $\pi/2$) acting on the
state of mode $j$. In case of pure squeezed vacua, the resulting
state can be expressed as
\begin{equation}
\label{state}
|\Psi(\phi)\rangle={\cal N}\left(|s_+\rangle|s_-\rangle+\mbox{e}^{\ii\phi}|s_-\rangle|s_+\rangle\right),
\end{equation}
where ${\cal N}=1/\sqrt{2[1+\mbox{sech} 2s\cos\phi]}$ is the
normalization factor and $|s_\pm\rangle=|\psi_{\pm s}\rangle$.
Evidently, being the superposition of two Gaussian states, the
ESV, $\ket{\Psi(\phi)}$, is not a Gaussian state
\cite{commentoGaussian}. It shows some resemblance with the
entangled coherent state~\cite{sanders1} $|\Phi_c\rangle={\cal
N}_c(|\alpha\rangle|-\alpha\rangle+\mbox{e}^{\ii\phi}|-\alpha\rangle|\alpha\rangle)$,
where $|\alpha\rangle=\mathrm{D}(\alpha)\ket{0}$ is a coherent state of amplitude
$\alpha$ and $\mathrm{D}(\alpha)=e^{(\alpha^*a-\alpha a^{\dag})}$ is the displacement  
operator~\cite{barnett}. The resemblance is due to the fact that
the overlap $\langle\alpha|-\alpha\rangle{\rightarrow}0$ as
$\alpha\rightarrow\infty$, making the components of the
superposition asymptotically orthogonal. Similarly, we have that
$\langle s_+| s_-\rangle\rightarrow 0$ as $s\rightarrow \infty$
thus making Eq.~(\ref{state}) the superposition of distinguishable
tensorial-product states. The task of our study is the
characterization and quantification of the entanglement shared
between a and b in states~(\ref{ESS}) and~(\ref{state}).

A well-known sufficient condition for inseparability is the
so-called Peres-Horodecki criterion~\cite{Peres} which is based on
the observation that the non-completely positive nature of the
partial transposition operation may turn an inseparable state into
a non-physical state. The signature of this non-physicality, and
thus of entanglement, is the appearance of a negative eigenvalue
in the eigenspectrum of the partially transposed density matrix of
a bipartite system. This approach cannot be directly applied to CV
states, spanning infinite dimensional Hilbert space. First
attempts to characterize the separability of CV states were
focused on second-order moments of some particular quadrature
operators~\cite{Simon,Duan}. For Gaussian states, whose
statistical properties are fully characterized by just
second-order moments, this criterion was proven to be necessary
and sufficient. Entanglement in non-Gaussian states, which are
characterized by the entire hierarchy of statistical moments, may
however fail to be detected. This is the reason why entanglement
tests for non-Gaussian states are based, in general, on
inequalities involving higher-order moments~\cite{Agarwal} or an
infinite hierarchy of inequalities~\cite{Shchukin,Hillery}. An
alternative to this kind of strategies is represented by the study
of the entangling power of the specific non-Gaussian state being
considered, {\it i.e.} the capacity of the state of inducing
entanglement (via just local interactions) in an initially
separable two-qubit system~\cite{Paternostro}.

In order to start our investigation, we first attempt to analyze
the pure state Eq.~(\ref{state}) under the point of view set by
the criteria introduced by Simon~\cite{Simon} and Duan {\it et
al.}~\cite{Duan}. As shown by Shchukin and Vogel~\cite{Shchukin},
both these criteria can be expressed in the following elegant
forms
\begin{eqnarray}
\label{SimonDuan}
C_{Simon} &=& \left|\begin{array}{ccccc}
1 & \langle a \rangle & \langle a^{\dag}\rangle & \langle b^{\dag} \rangle & \langle b\rangle \\
 \langle a^{\dag} \rangle& \langle a^{\dag 2} \rangle&\langle a^{\dag}a \rangle  & \langle a^{\dag}b^{\dag} \rangle & \langle a^{\dag}b \rangle   \\
 \langle a \rangle& \langle a^{2} \rangle & \langle a a^{\dag} \rangle & \langle ab^{\dag} \rangle &\langle ab \rangle  \\
\langle b \rangle & \langle ab \rangle & \langle a^{\dag}b \rangle &\langle b^{\dag}b \rangle  & \langle b^2 \rangle \\
 \langle b^{\dag} \rangle& \langle ab^{\dag} \rangle & \langle a^{\dag}b^{\dag} \rangle &\langle b^{\dag 2} \rangle  &  \langle bb^{\dag} \rangle
\end{array}
\right| \geq 0 \nonumber \\
C_{Duan} &=& \left| \begin{array}{ccc}
1 & \langle a \rangle & \langle b^{\dag} \rangle \\
\langle a^{\dag} \rangle & \langle a^{\dag}a \rangle &\langle a^{\dag}b^{\dag} \rangle  \\
\langle b \rangle & \langle ab \rangle & \langle b^{\dag}b \rangle
\end{array}
\right| \geq 0,
\end{eqnarray}
where $\langle{\cal O}\rangle$ stands for the expectation value of
the generic operator ${\cal O}$ calculated over the state being
investigated and each inequality is satisfied by separable states.
In the specific case of the state $\ket{\Psi(\phi)}$ in
Eq.~(\ref{state}), the mean of the arbitrary operator $AB$ (with
$A$ and $B$ general functions $A = A(a,a^{\dag})$, $B =
B(b,b^{\dag})$ of each mode's bosonic operators) is given by
\begin{equation}
\begin{split}
&\langle AB\rangle= \mathcal{N}^2\left(
\langle s_+ |A|s_+\rangle \langle s_- |B|s_-\rangle+\right.\\
&\left.\langle s_- |A|s_-\rangle \langle s_+ |B|s_+\rangle
+ 2 \mbox{Re}[e^{\ii\phi}\langle s_+ |A|s_-\rangle\langle s_- |B|s_+\rangle ] \right).
\end{split}
\end{equation}
By direct calculation, it is straightforward to conclude that all
the first- and second-order moments of $\ket{\Psi(\phi)}$ vanish,
with the exception of
\begin{equation}
\begin{split}
&\langle a^{\dag} a\rangle=2\mathcal{N}^2 \left(  \nu^2 - \frac{\cos(\phi)}{\cosh(2s)}[\nu^2 + \mu\nu\tanh(2s)] \right),\\
&\langle b^{\dag} b\rangle=\langle a^{\dag}a\rangle=\langle{a}{a^\dag}\rangle-1=\langle{b}{b^\dag}\rangle-1,\\
\end{split}
\end{equation}
where $\mu = \cosh s$ and $\nu = \sinh s$. It can be readily
checked, with these tools, that the criteria~(\ref{SimonDuan})
fail to prove inseparability even though the entanglement
properties of the state $\ket{\Psi(\phi)}$ are out of any doubts.
It is indeed possible to construct a discrete two-state basis,
following the recipe described in~\cite{iojacobmyung}, involving
the superposition of $\ket{s_+}$ and $\ket{s_-}$. This allows us
to reinterpret Eq.~(\ref{state}) as an effective entangled state
of two qubits whose entanglement depends, in general, on squeezing
parameter $s$ and relative phase $\phi$. In particular, for
$\phi=\pi$, the entanglement in the effective state is clearly
observable as soon as $s>0$.

However, this need not to be seen as a difficult setback since an
entangled coherent state~\cite{sanders1} also remains undetected
in this manner~\cite{Agarwal}. We simply have to look for a more
powerful test. A sufficient criterion based on higher-order
moments was proposed by Shchukin and Vogel~\cite{Shchukin} who
observed a useful feature of the determinant of the
infinite-dimensional matrix $M$ composed of the statistical
moments
\begin{equation}
\begin{split}
M_{ij}(\rho^{PT})&=\langle a^{\dag i_1}a^{i_2}b^{\dag i_3}b^{i_4} a^{\dag j_2}a^{j_1}b^{\dag j_4}b^{j_3}\rangle^{PT}\\
&=\langle a^{\dag i_1}a^{i_2}b^{\dag j_3}b^{j_4} a^{\dag j_2}a^{j_1}b^{\dag i_4}b^{i_3}\rangle,
\end{split}
\end{equation}
where $PT$ stands for partial transposition while $i,j$ are the multi-indices $i = (i_1,i_2,i_3,i_4)$ and $j = (j_1,j_2,j_3,j_4)$. The ordering of the multi-indices is such that $i>j$ iff
\begin{eqnarray}
 &|i|>|j|& ~ \mbox{or}  \\
 &|i|=|j|& ~
 \mbox{and} ~\exists k:i_k>j_k ~\mbox{while}~ i_l = j_l, \forall l >k,   \nonumber
\end{eqnarray}
with $|i|=i_1+i_2+i_3+i_4$ (analogously for $|j|$). Now, the state
characterized by a density matrix $\rho$ is inseparable ({\it
i.e.} has negative partial transpose), if there exists a leading
principal minor (or just any principal minor, as was stressed
in~\cite{Miranowicz}) for which the relevant determinant is
negative. More formally, $\rho$ is inseparable if
\begin{equation}
\begin{split}
\exists N = 1,2,\cdots&~\mbox{and}~r = (r_1,r_2,\ldots,r_N):\nonumber \\
 &\mbox{det} M^r(\rho^{PT})<0,
\end{split}
\end{equation}
where $1\leq r_1<r_2<\cdots<r_N$ and $M^r(\rho^{PT})$ is obtained
from $M(\rho^{PT})$ by deleting all rows and columns except those
labelled by $r_1,{\hdots},r_N$. This formalism also includes other
separability criteria. Simon's criterion~\cite{Simon}, for
instance, is obtained by setting $r_{Simon}=(1,2,3,4,5)$ and Duan
{\it et al.}'s~\cite{Duan} by $r_{Duan} = (1,2,4)$. Other criteria
can be obtained in a similar way.

The important task in devising a separability criterion for
state~(\ref{state}) is therefore finding the right vector $r$ and
considering an appropriate determinant. It turns out that this can
indeed be done and the full criterion can be given in the form of
the following inequality
\begin{equation}
\label{petr}
\left|
\begin{array}{ccccc}
1 & \langle a^{\dag}b\rangle  &\langle a^{\dag}b^{\dag} \rangle &\langle  ab \rangle & \langle ab^{\dag} \rangle \\
\langle ab^{\dag}\rangle &\langle  aa^{\dag}bb^{\dag}\rangle &\langle aa^{\dag}b^{\dag 2}\rangle  &\langle a^2 bb^{\dag}\rangle  &\langle a^2 b^{\dag 2}\rangle \\
\langle ab \rangle &\langle  aa^{\dag} b^2\rangle &\langle  aa^{\dag}b^{\dag}b\rangle &\langle  a^2b^2\rangle & \langle a^2b^{\dag}b\rangle \\
\langle a^{\dag}b^{\dag}\rangle & \langle a^{\dag 2} bb^{\dag}\rangle & \langle a^{\dag 2} b^{\dag 2}\rangle & \langle a^{\dag}a bb^{\dag}\rangle  & \langle a^{\dag}ab^{\dag 2}\rangle  \\
\langle a^{\dag}b\rangle & \langle a^{\dag 2}b^2\rangle & \langle a^{\dag 2}b^{\dag}b\rangle & \langle a^{\dag}ab^2\rangle &\langle a^{\dag}ab^{\dag}b \rangle
\end{array}
\right| <0.
\end{equation}
This condition, implying negativity of partial transposition and entanglement, is satisfied for any value of $\phi$ and $s$, although for low values of initial squeezing the violation may be difficult to observe. We have thus designed a proper condition for testing the quantum correlations shared between modes a and b in the non-Gaussian state $\ket{\Psi(\phi)}$.







As already stated, the use of inequalities based on hierarchies of
statistical moments is not the only way we have to infer
entanglement in a bipartite non-Gaussian state. Alternatively (and
somehow more operatively) we can also check for the capability of
the state at hand to induce entanglement, by means of just
bi-local interactions, into an initially separable state of two
non-interacting qubits. This has been formalized by the so-called
entangling power test, which has been applied in order to get some
physical insight into some problems of quantum state
characterization and has demonstrated to be powerful under many
points of view~\cite{Paternostro,iojacobmyung,Paternostro2}. A
full description of the protocol is certainly not the main focus
of this work and we refer to the available literature for a
detailed analysis of the entangling power test. Here, we just
sketch the main points related to the protocol.

\begin{figure}[t]
{\bf (a)}\hskip3.5cm{\bf (b)}
\centerline{\psfig{figure=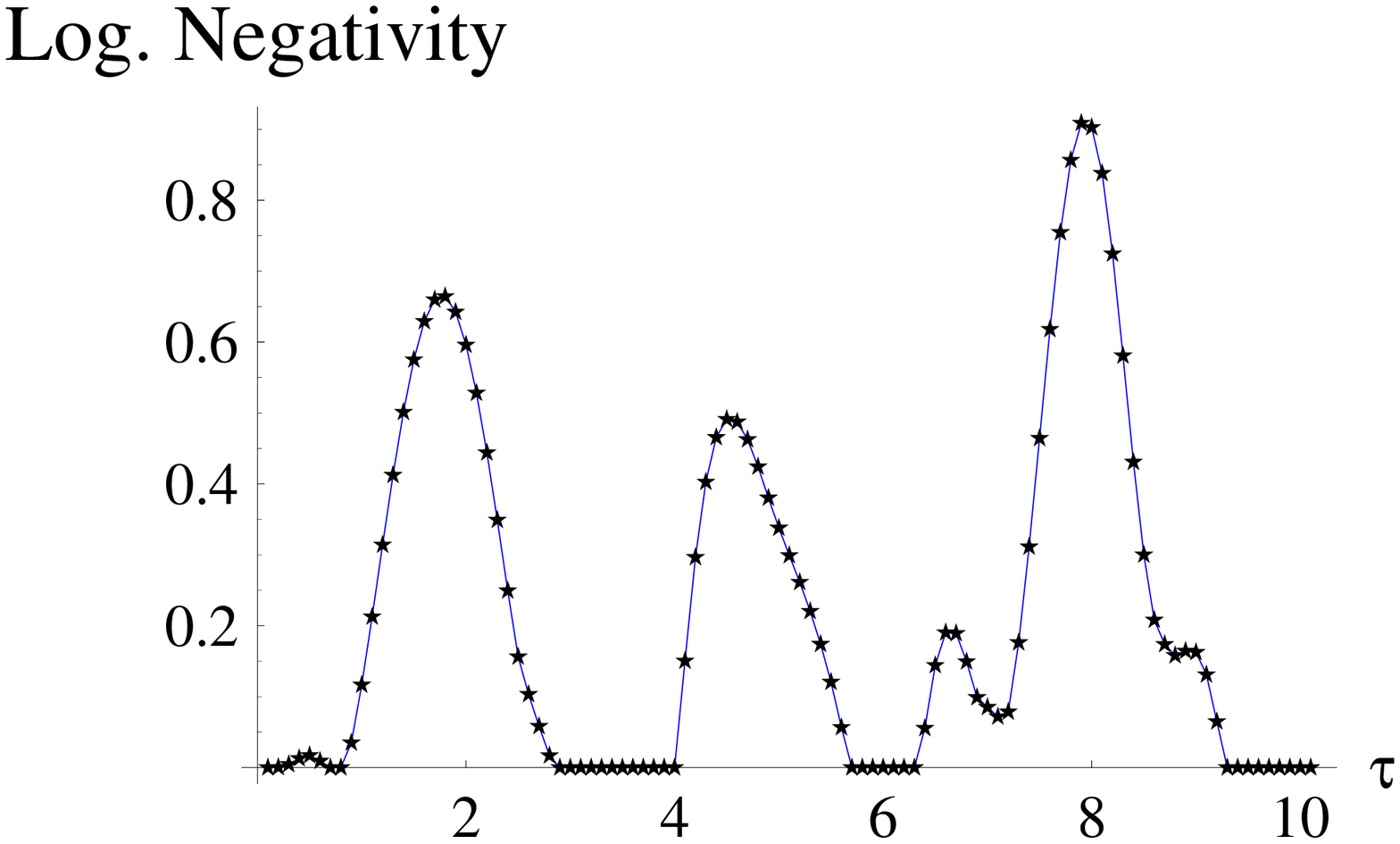,width=4.8cm,height=3.0cm}\hskip-0.5cm
\psfig{figure=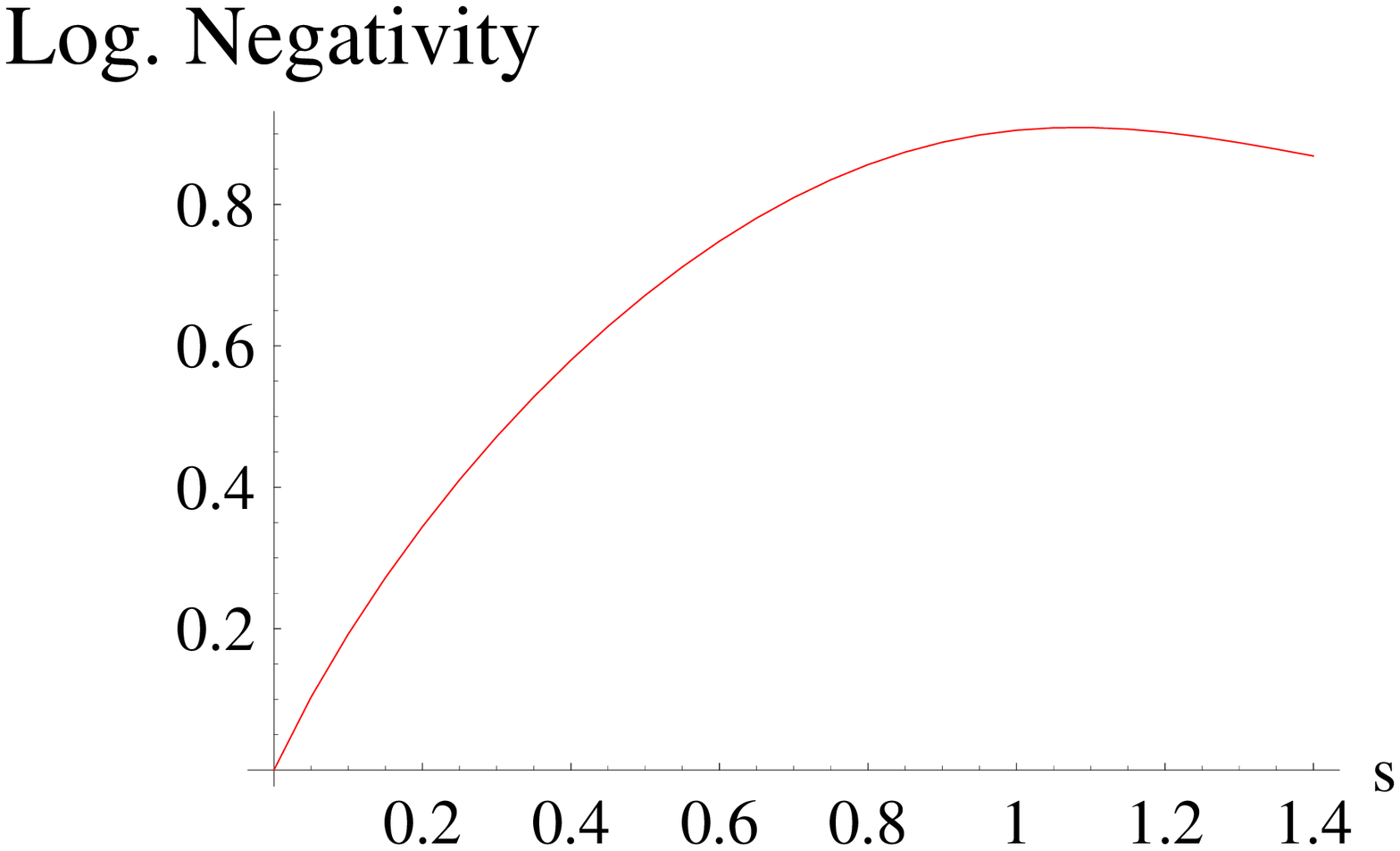,width=4.3cm,height=3.cm}} \caption{(Color online) {\bf
(a)}: Entangling power as a function of the (dimensionless)
rescaled interaction time $\tau$, for $\phi=0$  and $s=1.1$. {\bf
(b)}: Logarithmic negativity against $s$, for $\tau=8$.}
\label{entpower}
\end{figure}

We assume that each mode participating to state (\ref{state})
interacts for a time $t$ with a qubit initially prepared in its
ground state. Each interaction is local to a qubit-mode subsystem.
In the spirit of~\cite{Paternostro,iojacobmyung,Paternostro2},
here we assume a resonant Jaynes-Cummings coupling of equal
strength $g$ for each local interaction. The state of the two CV
modes is then traced out in order to get the reduced (mixed)
density matrix for the qubits. The Peres-Horodecki criterion is
then applied to the qubit state. Obviously, if any entanglement is
found between the qubits, this can only mean that modes a and b
were originally entangled, as the local interactions alone can not
set entanglement between the qubits. At the same time, the amount
of entanglement, here quantified by the logarithmic
negativity~\cite{logneg}, represents a lower bound to the
entanglement shared between the modes~\cite{iojacobmyung}. The
results are shown in Fig.~\ref{entpower} {\bf (a)} against the
rescaled interaction time $\tau=gt$, for $\phi=0$ and $s=1.1$. A
qualitatively analogous picture could be given for the case of
arbitrary values of $\phi$ in Eq.~(\ref{state}).
A large amount of transferred entanglement ($>0.9$) is found to
occur at $\tau\simeq{8}$ for this choice of $s$, which turns out
to be the optimal value for entanglement transfer. The optimality
of $s=1.1$ is demonstrated in Fig.~\ref{entpower} {\bf (b)}, where
the logarithmic negativity is plotted against $s$, for $\tau=8$.
We have also checked that, for $\tau\le{10}$, the transferred
entanglement corresponding to other choices of $s$ and other
rescaled interaction times is always smaller than the value
achieved at $\tau=8$ with $s=1.1$. The considerations of
experimental observability of the entangling power discussed
in~\cite{Paternostro,Paternostro2,iojacobmyung} are valid in the
context of the present work.

Given the similarity between entangled coherent states and the
class of ESV's, it is not difficult to think about applications of
ESS's in QIP schemes such as those described in~\cite{noinjp}. The
large entanglement content (quantitatively demonstrated in the
next Subsection), corresponding to specific choices of $\phi$,
encourages this conjecture. In Section~\ref{teleswap} we
investigate their performances as a quantum channel in some
paradigmatic protocols of QIP.

Let us now briefly discuss a generalization of the state~(\ref{state}). As a natural extension we consider the state
\begin{equation}\label{generalized}
|\Psi'\rangle = \mathcal{N}'\left(|\alpha_+,\beta_-\rangle + \ee^{\ii\phi}
|\beta_-,\alpha_+\rangle\right),
\end{equation} 
where $\mathcal{N}'$ is the normalization factor and 
\begin{eqnarray}
|\alpha_+\rangle =\mathrm{D}(\alpha)|s_+\rangle,\ |\beta_-\rangle =  \mathrm{D}(\beta)|s_-\rangle
\end{eqnarray} 
are displaced squeezed states. Eq.~(\ref{generalized}) can be seen as a conjuncture of entangled squeezed states and entangled coherent states~\cite{sanders1}. The detailed discussion of the entanglement of such a state is beyond the scope of the present work. However, it is worth discussing some qualitative features which are useful in order to understand the pattern of quantum correlations in Eq.~(\ref{generalized}). The overlap $|\langle\alpha_+|\beta_-\rangle|^2$ is critical in determining the entanglement of $\ket{\Psi'}$, which increases as the overlap tends to zero. Quantitatively
\begin{equation}
\label{over}
\bigl|\langle \alpha_+|\beta_-\rangle\bigr|^2 = \frac{1}{\cosh{(2r)}} \exp\left(-\frac{|\beta-\alpha|^2}{\cosh(2r)}\right),
\end{equation} 
which is a monotonically decreasing function of the difference $|\beta-\alpha|$ between the single-mode displacements. The dependence from the squeezing factor $r$, on the other hand, is such that the overlap~(\ref{over}) is maximized at $r=\frac{1}{2}\sec\!{\mathrm h}^{-1}(|\beta-\alpha|^{-2})$. Therefore, there are situations in which 
an increase of the squeezing can actually lead to the enlargement of the overlap and consequently to the reduction of the entanglement. Note that, in general, state~(\ref{generalized}) cannot be created by the transformation introduced in Eq.~(\ref{Transformation}) and an {\it ad hoc} approach theory should be developed. Further generalizations can be achieved by considering the states $\hat{S}(\ee^{i\xi_1}|s_1|)|0\rangle$ and $\hat{S}(\ee^{\ii\xi_2}|s_2|)|0\rangle$, which are squeezed vacua of arbitrary phases, instead of $|s_\pm\rangle$.

\subsection{Quantifying the entanglement in a mixed ESS}
Having established the existence of entanglement of
state~(\ref{state}) we now concentrate on its quantification with
respect to pure and mixed ESS's. In general, we will deal with
mixed non-Gaussian states for which many proposed entanglement
monotones may be proven too difficult to compute and, therefore,
unsuitable for our purely demonstrational aims. For this reason,
in what follows we will focus on the use of the entanglement
measure based on logarithmic negativity, which is computable as
\begin{equation}
LN(\rho) = \log_2 \Vert \rho^{PT}\Vert,
\end{equation}
where the trace norm $\Vert A \Vert = \mbox{Tr}(\sqrt{A^{\dag}A})$
(with $\mbox{Tr}$ the trace of a matrix) is calculated as the sum
of the absolute values of the eigenvalues of $A$. These can be
numerically calculated by truncating the dimension of the
infinite-dimensional Hilbert space where the ESV is defined to a
sufficiently large (but finite) computational subspace.

However, as a consistency check, let us first take advantage of
the purity of Eq.~(\ref{state}) and calculate the entanglement of
formation~\cite{EOF}
\begin{equation}
E_F = -\mbox{Tr}[\rho_A\log_2\rho_A],
\end{equation}
where $\rho_A = \mbox{Tr}_B[\rho]$.
\begin{figure}
\centerline{\psfig{figure=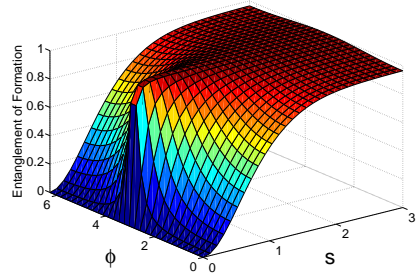,width=0.7\linewidth}}
\caption{\label{EoF} (Color online) Entanglement of formation with respect to squeezing parameter $s$ and relative phase $\phi$.}
\end{figure}
As can be seen in Fig.~\ref{EoF}, entanglement is present for all
values of $s$ and $\phi$. For $s\gg{1}$, $E_F$ approaches a full
ebit, which is clearly the manifestation of a vanishing overlap
$\langle s_+|s_-\rangle$ (quantitatively, $\langle
s_+|s_-\rangle\simeq10^{-2}$, for $s=5$). The corresponding
effective two-qubit state obtained by using the discretized
computational basis (see Subsection~\ref{uno}) is then maximally
entangled. However, for $\phi =\pi$, $E_F$ approaches a full ebit
even for small amount of initial squeezing. This behavior can be
easily explained as follows. Let us consider the following
single-mode states, which are superpositions of states squeezed in
two opposite directions
\begin{equation}
\label{q-s}
\begin{split}
|\varphi_+\rangle&={\cal N_{\varphi_{+}}}(|s_+\rangle+|s_-\rangle),\\
|\varphi_-\rangle&={\cal N_{\varphi_{-}}}(|s_+\rangle-|s_-\rangle),
\end{split}
\end{equation}
where ${\cal N_{\varphi_{\pm}}}=\left[\frac{\sqrt{\cosh
s}}{2(\sqrt{\cosh s}\pm1)}\right]^{1/2}$ are normalization
factors. A squeezed vacuum can be represented as a superposition
of Fock states as $|\psi_s\rangle=\frac{1}{\sqrt{\cosh
s}}\sum_{n=0}^{\infty}\frac{\sqrt{(2n)!}}{n!}\left(-{\frac{1}{2}}\tanh
s \right)^n|2n\rangle$~\cite{loudon}. With this decomposition, the
superposition states take the forms
\begin{equation}
\begin{aligned}
|\varphi_+\rangle&={\cal N}_+\sum_{k=0}^\infty \frac{\sqrt{(4k)!}}{(2k)!}\left({\frac{1}{2}}\tanh s\right)^{2k}|4k\rangle,\\
|\varphi_-\rangle&={\cal N}_-\sum_{k=0}^\infty \frac{\sqrt{(4k+2)!}}{(2k+1)!}\left({\frac{1}{2}}\tanh s\right)^{2k+1}|4k+2\rangle.
\label{q-s-1}
\end{aligned}
\end{equation}
These are quadro-multiple states which have been discussed by
Sanders for the study of superpositions of squeezed
states~\cite{sanders}. It is obvious that the two states are
orthogonal to each other. Therefore, after normalization, the
state $\ket{\chi}\propto|\varphi_+\rangle|\varphi_-\rangle \pm
|\varphi_-\rangle|\varphi_+\rangle$ should bear an ebit. By
putting their definitions (\ref{q-s}) into $\ket{\chi}$, we
realize that the states carrying one ebit are exactly the ESV
states with $\phi=\pi$. Note, that very similar reasoning would be
applied explaining the entanglement of the state $|\Phi(\pi)
\rangle \propto |s_+\rangle|s_+\rangle - |s_-\rangle|s_-\rangle$.
This is not surprising as such a state can be  obtained from the
$|\Psi(\pi)\rangle$ state by a local $\pi/2$ phase shift applied
to one of the modes.

Let us return to the case of the mixed entangled squeezed state
(\ref{ESS}) and discuss the influence of imperfection on the
entanglement. As a starting point we model a mixed squeezed state
by using two types of noise affecting an initial pure squeezed
vacuum.
\begin{figure}[b]
\centerline{\psfig{figure=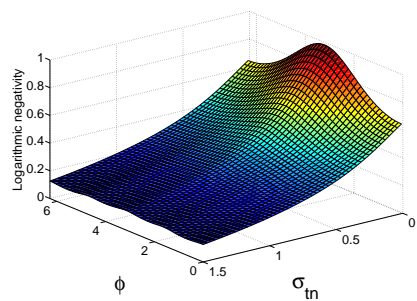,width=0.7\linewidth}}
\caption{(Color online) Logarithmic negativity of entangled state of squeezed
states with thermal noise, against the relative phase $\phi$ and
the width of the Gaussian weight $\sigma$. The initial squeezing
considered is $s=1$.} \label{LNthermal}
\end{figure}
The variances of the squeezed (sq) and anti-squeezed (as)
quadratures of a squeezed vacuum state affected by thermal
Gaussian noise can be written in the form
\begin{equation}
V_{sq} = \frac{1}{2}\ee^{-2s} +\sigma_{tn} \quad V_{as} = \frac{1}{2}\ee^{2s} + \sigma_{tn},
\end{equation}
where $\sigma_{tn}$ characterizes the phase-insensitive Gaussian
noise. The density matrix of such a state can be obtained as the
classical mixture of randomly displaced squeezed states weighted
by a Gaussian distribution of width $\sigma_{tn}$ as
\begin{equation}
\label{thermalnoise}
\rho_{T} = \int\frac{\ee^{-|\alpha|^2/\sigma_{tn}}}{\pi\sigma_{tn}} {\mathrm D}(\alpha) \rho_{s+} {\mathrm D}^{\dag}(\alpha)  \dd^2 \alpha.
\end{equation}
\begin{figure}[h]
\centerline{\psfig{figure=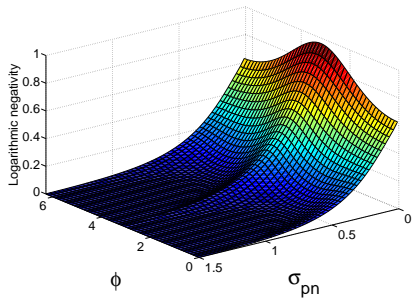,width=0.7\linewidth}}
\caption{(Color online) Logarithmic negativity of  ESV affected by
phase-diffusion noise, against the relative phase $\phi$ and the
width of the Gaussian weight $\sigma$. The initial squeezing
considered is $s=1$.} \label{LNphase}
\end{figure}
The second realistic type of noise we consider is caused by a
randomly fluctuating refractive index of the medium through which
the initial signal passes in order to generate the ESV's, leading
to the density matrix having the form
\begin{equation}
\label{phasenoise}
\rho_{P} = \int\frac{\ee^{-\varphi^2/2\sigma_{pn}}}{\sqrt{2\pi\sigma_{pn}}} \mathrm{R}(\varphi) \rho_{s+} \mathrm{R}^{\dag}(\varphi) \dd \varphi.
\end{equation}
where $\mathrm{R}(\varphi)=\exp(\ii\varphi{a}^{\dag}a)$ is the phase-shift
operator already introduced and we have considered Gaussian
fluctuation of the refractive index. The logarithmic negativity
corresponding to states~(\ref{thermalnoise})
and~(\ref{phasenoise}) is shown in Figs.~\ref{LNthermal}
and~\ref{LNphase}, where the entanglement is studied against the
relative phase $\phi$ and the spread of the Gaussian distributions
$\sigma$. Evidently, as soon as the spread in the Gaussian weight
increases, the entanglement diminishes. However, while the
logarithmic negativity becomes zero for most of the values of
$\phi$ for $\sigma_{pn}\ge{1}$ in the case of phase noise, it is
still as large as $0.1$ for $\sigma_{tn}\simeq{2}$, uniformly with
respect to $\phi$. An analogous decrease in the entanglement can
be found by considering again the entangling power test. In this
case, in order to simplify the (otherwise computationally
prohibitive) calculations, the mixedness of the ESV is
artificially modelled as follows. We consider a single-mode
squeezed state $\ket{\psi_s}$ and an ancillary vacuum state mixed
at a beam splitter (BS) of transmissivity $\cos^2\theta$. By
tracing out the output ancillary state, we get an output mixed
squeezed state whose purity depends on the transmissivity of the
BS. This models each of the $\rho_{s+,j}$ entering
Eq.~(\ref{ESS}). By applying the entangling power test to the
resulting state, we have quantitatively found a transferred
entanglement no larger than $0.4$ for the optimal squeezing
$s=1.1$ and $\phi=0$.

\section{Entanglement swapping and teleportation}
\label{teleswap}
In this Section we briefly discuss the possibility of implementing
some basic protocols for QIP by using the class of ESS's. In
particular, as a prototypical example where the entanglement of a
channel plays a crucial role, we address the probabilistic
realization of entanglement swapping. We also show how a quantum
teleportation protocol can be realized.

Starting with two entangled subsystems embodied by the two pairs
of modes 1 and 2 and 3 and 4, the aim of entanglement swapping
is to create an entangled state of modes 1 and 4. This can be
performed by a joint measurement of modes 2 and 3 and either
post-selection or local operations performed on remaining
modes~\cite{swapping}. Let us consider the initial states for the
two subsystems to be the tensorial product of the two ESV's
\begin{equation}
\label{swapp}
\begin{split}
|\Psi(\pi)\rangle_{12}&=\mathcal{N}(|s_+\rangle|s_-\rangle-|s_-\rangle|s_+\rangle)_{12},\\
|\Phi(\pi)\rangle_{34}&=\mathcal{N}(|s_+\rangle|s_+\rangle-|s_-\rangle|s_-\rangle)_{34}.
\end{split}
\end{equation}
As stated above, the task is to create one ebit of entanglement
within the subsystem 1 and 4. In order to achieve this, the modes
$2$ and $3$ are superimposed on a balanced BS realizing the
transformations $a_2\rightarrow(a_2+a_3)/\sqrt{2}$ and
$a_3\rightarrow(a_3-a_2)/\sqrt{2}$. Consequently, the initial
state $|\Psi(\pi)\rangle_{12}\otimes|\Phi(\pi)\rangle_{34}$
becomes
\begin{equation}
\label{ES}
\begin{split}
&\mathcal{N}^2|\psi\rangle_{1234}=\\
&\quad\quad\ S_{23}(s)|0,0\rangle_{23}|s_-,s_-\rangle_{14} \\
&\quad +S_2(-s)S_3(-s)|0,0\rangle_{23}|s_+,s_-\rangle_{14}\\
&\quad +S_2(s)S_3(s)|0,0\rangle_{23}|s_-,s_+\rangle_{14}\\
&\quad +S_{23}(-s)|0,0\rangle_{23}|s_+,s_+\rangle_{14},
\end{split}
\end{equation}
where $S_{ab}(s)|0,0\rangle_{ab}=({\cosh
s})^{-1}\sum_{n=0}^{\infty}(\tanh{s})^n|n,n\rangle_{ab}$ is a
two-mode squeezed vacuum state. The second and third terms in
Eq.~(\ref{ES}) contain only even number of photons in modes $2$
and $3$ while the first and last terms contain both even and odd
numbers. The entanglement swapping can now be achieved by
performing the projection onto odd photon-number states of modes
$2$ and $3$ formally described by $\Pi_{23}=\sum_k
|2k+1\rangle_{23}\langle{2}k+1|$ and tracing over these modes.
This leads to the output state
\begin{equation}
\left[\mathcal{N}^2\sum_{k=0}^{\infty}\frac{(\tanh s)^{2 + 4k}}{(\cosh{s})^2 }\right]|\Phi(\pi)\rangle_{14}\langle \Phi(\pi)|.
\end{equation}
The desired entangled state of modes $1$ and $4$ is obtained with
a probability which can be easily evaluated to be $1/4$.  A
probabilistic entanglement swapping is possible by using two
entangled resources embodied in ESV's.

A similar approach can be used in order to realize quantum
teleportation~\cite{teleportation}. Let us assume that our task is
the teleportation between two spatially separated nodes $A$ and
$B$ of an unknown state of an input mode $I$
$(\alpha|s_+\rangle+\beta|s_-\rangle)_{I}$. We consider the state
$|\Phi(\pi)\rangle_{12}$ as an entangled resource shared by $A$
and $B$. At the local node $A$, the input mode $I$ and mode $1$ of
the entangled state are mixed on a BS and a projection analogous
to the one described in the entanglement swapping protocol is
performed. By projecting onto odd photon-number states of the
subsystem $I$ and 1, a classical message is sent to node $B$ which
holds mode $2$. At this stage, a $\pi/2$ rotation in phase space
on mode $2$ allows for the reconstruction of the unknown input
state at node $B$. Analogously to the entanglement swapping, the
entire procedure succeeds with a probability equal to $1/4$.

\section{Generation of ESS}
\label{scheme}
\begin{figure}[ht]
\centerline{\psfig{figure=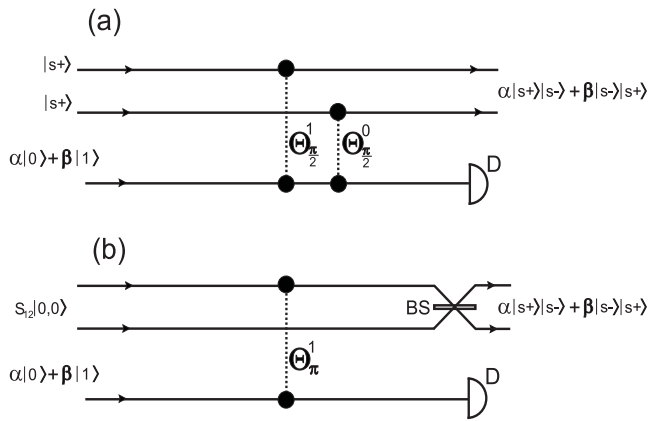,width=8.5cm,height=5.5cm}}
\caption{\label{setup} Generation schemes for ESS's. The vertical
dotted lines represent the controlled phase-flip operations, D
stands for a detector measuring in the $|\pm\rangle$ basis and BS
is a balanced beam splitter.}
\end{figure}

We complete our characterization of ESS's by briefly addressing
the problem of generating this class of states. General recipes
can be given for the realization of an ESS, both exploiting
similar physical mechanisms. In details, in order to create ESS's
we require the evolution described by the operator ${\cal T}$.
This can be accomplished in several ways, two of them depicted on
Figs.~\ref{setup}, both requiring coupling of two CV systems with
an ancillary two-level system. In both the schemes, the squeezed
states need not to be pure, although we are going to assume this
quality in order to simplify the description. The most
straightforward way employs a pair of identical single mode
squeezed states that are coupled to the ancillary system via
controlled phase-flip operations acting as
\begin{equation}
\begin{split}
\Theta^0_{\pi/2}|s_+\rangle|0\rangle&= |s_-\rangle|0\rangle,\quad \Theta^0_{\pi/2}|s_+\rangle|1\rangle = |s_+\rangle|1\rangle,\\
\Theta^1_{\pi/2}|s_+\rangle|0\rangle&= |s_+\rangle|0\rangle,\quad \Theta^1_{\pi/2}|s_+\rangle|1\rangle = |s_-\rangle|1\rangle,
\end{split}
\end{equation}
where $|0\rangle$ and $|1\rangle$ are the states of the
computational basis of the ancillary qubit (cf. Fig.~\ref{setup}
{\bf (a)})~\cite{jeong}.

The second possibility is to use a two-mode squeezed vacuum
$S_{12}$ as a resource. In this case, it is sufficient to perform
a single controlled phase-shift, by an amount of $\pi$, on the
joint state of an ancilla and one of modes $j=1,2$, followed by a
mixing of modes $1$ and $2$ using a balanced BS (cf.
Fig.~\ref{setup} {\bf (b)}). Notice however that in the first
scheme we do not need to prepare the input modes in independent
squeezed states. They can be obtained by using a two-mode squeezed
state whose components are previously mixed at a balanced BS. In
this way, a single squeezed resource is exhausted in the
generation of an ESS, in both the schemes.

Given an arbitrary ancillary state of the form $\alpha|0\rangle +
\beta|1\rangle$ ($|\alpha|^2+|\beta|^2 = 1$), both these
possibilities lead to a joint state of the three participating
modes reading
\begin{equation}
\alpha|s_+\rangle|s_-\rangle|0\rangle + \beta|s_-\rangle|s_+\rangle|1\rangle.
\end{equation}
The ancillary quibit is now measured in the rotated basis
$|+\rangle = (|0\rangle+|1\rangle)/\sqrt{2}$,\,$|-\rangle
=(|0\rangle-|1\rangle)/\sqrt{2}$ and, conditioned on the outcome
of the measurement, the resulting two-mode CV states read
\begin{equation}
\alpha|s_+\rangle|s_-\rangle \pm \beta|s_-\rangle|s_+\rangle.
\end{equation}
The $+$ ($-$) sign corresponds to the ancillary state $\ket{+}$
($\ket{-}$) being measured. The generality of the described
protocols allows for practical implementations in various physical
setups. 

The key part in the discussed setups is the cross-Kerr type interaction $\Theta_{\gamma}=e^{\ii \gamma a^{\dag}a b^{\dag} b}$ of the ancilla qubit with the $j$-th mode of interest. This interaction plays a critical  
role in the engineering of non-classical CV states like superpositions of coherent states and entangled coherent states and in quantum computation with coherent states~\cite{jeong,Paternostro3}. The generation of ESS's requires either two cross-Kerr interactions with parameters $\gamma = \pi/2$ (for the case of the scheme presented in Fig.~\ref{setup} {\bf (a)}), or a single one with $\gamma = \pi$ (Fig.~\ref{setup} {\bf (b)}). It can be easily checked that in both cases, as a result of the photon-distribution properties of squeezed states, the required effective rate of nonlinearity is half the analogous nonlinearity needed for the generation of entangled coherent states (by using the setup in Figs.~\ref{setup}).

In a setup of cavity-quantum electrodynamics (cavity-QED), the scheme in Fig.~\ref{setup} {\bf (a)} can be realized with the ancillary qubit being embodied by a flying two-level atom, which passes through a microwave cavity supporting a single-mode field~\cite{commentosqueezing}. If the atom-field interaction is set to be far off-resonant (with a large detuning being set by the ac Stark shift effect on the atomic level spacing induced by a static intracavity electric field), the effective time-evolution operator regulationg the dynamics of the joint atom-field state, is of the required cross-Kerr type. Due to the long lifetimes of both the flying qubit and the cavity field~\cite{Haroche}, the interaction time leading to $\gamma\sim{\pi}$ can be achieved with current technology~\cite{Haroche,noinjp}. On the other hand, still within cavity-QED, the required coupling responsible for the conditioned phase-flip in Fig.~\ref{setup} {\bf (b)} can be accomplished by arranging an interaction between one mode of the two-mode squeezed state with a single atom trapped inside an optical cavity, as in the scheme described by Wang and Duan~\cite{Wang}. In this case, one can take an advantage of the recently achieved regime of strong interaction between the cavity field and the trapped atom where the conditioned phase-flip can be obtained within the coherence time of both the atom and the cavity (see Wang and Duan in~\cite{Wang} for a detailed discussion).

An all-optical implementation scheme would require a path encoded photonic qubit as an ancilla. This can be straightforwardly achieved using the polarization of a photon and a polarization beam splitter.  
The cross-Kerr coupling could then be implemented in doped optical fibres. However, the strength of Kerr nonlinearity is very low and makes the approach unrealistic for the current state of the art~\cite{fibrebad}. A more promising approach relies on the use of double-electromagnetically induced transparency (double-EIT)~\cite{OL}. Here, the ancillary light field and the target mode are present in an effective highly-nonlinear medium where the group velocities of the two fields are simultaneously reduced (by the EIT mechanism).  If the effectively nonlinear medium is characterized by a sufficient degree of symmetry, the two velocities will be of the same (small) value. Consequently, the interaction time between two light fields can be increased in a way to achieve $\gamma\sim{\pi}$. In these years, various energy configurations of the effective nonlinear medium and many coupling schemes with light have been considered, in order to achieve a perfect double-EIT, free from residual absorption (which limits the achivable nonlinear coupling rate), Doppler-broadening (which spoils the transparency window inducd by EIT) and dephasing.  The desired phase shift can be then tuned by a proper choice of the strength and length of the signal pulse. With current technology, a single photon induced phase shift of $\approx\pi/12$ can be expected \cite{Chen}.

\section{Remarks}
\label{conclusions} 
In the important context of quantum state
engineering and characterization, we have studied the entanglement
properties of the special class of ESS's. The bottleneck set by
the intrinsic non-Gaussianity of the states considered is bypassed
by considering an entanglement criterion based on higher-order
statistical moments and by the application of the entangling power
test. Clear physical interpretations for the entanglement found
for different values of the relative phase in an ESS have been
provided. This helps the comprehension of the quantitative results
achieved by the use of the logarithmic negativity. As examples for
applications of an ESS, we have suggested two probabilistic
schemes for entanglement swapping and teleportation. Finally, we
have briefly described a general generation scheme based on the
use of a single ancilla and non-linear interactions whose
practical implementation can be found in various physical setups.

\acknowledgments We acknowledge financial support from the UK
EPSRC, the European Social Foundation fund and the Leverhulme
Trust (ECF/40157). 


\begin{thebibliography}{99}
\bibitem{commentoGaussian} 
The characteristic function of $\ket{\Psi(\phi)}$ will be the sum of Gaussian functions, which is not Gaussian any more.
\bibitem{commentosqueezing} Single-mode squeezed states are routinely generated in the microwave regime by superconducting Josephson parametric amplifiers and coupled to the microwave cavity through waveguides.
\end{thebibliography}
\end{document}